\documentclass[letterpaper,floatfix,reprint,aps,prx,superscriptaddress,onecolumn]{revtex4-2}

\usepackage{graphicx}
\usepackage{gensymb}
\usepackage{tabularx}
\usepackage{color}
\usepackage{amsmath}
\usepackage{hyperref}
\usepackage{xcolor}
\usepackage{placeins, float}

\definecolor{gcb4}{HTML}{33A02C}

\preprint{APS/123-QED}

\begin{document}

\title{Engineering morphogenesis of cell clusters with differentiable programming}

\author{Ramya Deshpande}
\thanks{These authors contributed equally to this work.}
\affiliation{School of Engineering and Applied Sciences, Harvard University, Cambridge MA 02138}

\author{Francesco Mottes} 
\thanks{These authors contributed equally to this work.}
\affiliation{School of Engineering and Applied Sciences, Harvard University, Cambridge MA 02138}
\affiliation{Department of Computational Biology, University of Lausanne, Lausanne, Switzerland}

\author{Ariana-Dalia Vlad}
\affiliation{Harvard College, Cambridge, MA 02138}
\author{Michael P. Brenner}
\affiliation{School of Engineering and Applied Sciences, Harvard University, Cambridge MA 02138}
\author{Alma dal Co}
\affiliation{Department of Computational Biology, University of Lausanne, Lausanne, Switzerland}

\date{\today}

\begin{abstract}

Understanding the rules underlying organismal development is a major unsolved problem in biology. Each cell in a developing organism responds to signals in its local environment by dividing, excreting, consuming, or reorganizing, yet how these individual actions coordinate over a macroscopic number of cells to grow complex structures with exquisite functionality is unknown.  
Here we use recent advances in automatic differentiation to discover local interaction rules and genetic networks that yield emergent, systems-level characteristics in a model of development. We consider a growing tissue with cellular interactions mediated by morphogen diffusion, cell adhesion and mechanical stress. Each cell has an internal genetic network that is used to make decisions based on the cell's local environment. We show that one can learn the parameters governing cell interactions in the form of  interpretable genetic networks for complex developmental scenarios, including directed axial elongation, cell type homeostasis via chemical signaling and homogenization of growth via mechanical stress. When combined with recent experimental advances measuring spatio-temporal dynamics and gene expression of cells in a growing tissue, the methodology outlined here offers a promising path to unraveling the cellular bases of development.

\end{abstract}

\maketitle

Morphogenesis, the emergence of distinct anatomical forms from dividing cellular assemblies, hinges on the delicate coordination of diverse cellular and molecular processes across multiple spatial and temporal scales. Programming cells to achieve desired morphogenetic outcomes  is extremely challenging. Unlike systems guided by centralized control signals \cite{rubenstein2014programmable}, most developmental complexity arises from the interplay of local communications, sensing, and information processing within individual cells, with minimal external influence \cite{collinet_programmed_2021}. The emergent nature of collective cell phenomena, coupled with the complexity of modeling them in adequate detail, imposes severe limits to the effectiveness of computational approaches to inverse design.

Recent experimental advances offer unprecedented opportunities for morphogenetic engineering \cite{davies_using_2017,velazquez_programming_2018}, enabling the engineering of tailored cellular interactions, with applications ranging from population control \cite{ma_synthetic_2022} to programming multicellular assembly patterns \cite{toda_engineering_2020}. Leveraging the intrinsic self-organizing tendencies of cell clusters has also facilitated the production of diverse organoids that mimic multiple aspects of real tissues and organs  \cite{lancaster_guided_2017,lewis_self-organization_2021,brassard_engineering_2019}. The ability to quantitatively guide existing self-organizing developmental programs or engineer novel ones holds significant implications for both fundamental scientific understanding and biotechnology \cite{sthijns_building_2019, werner_self-organization_2017}. Many of these successes, however, heavily rely on manually crafted and qualitative cellular interaction rules and protocols. Heuristic experimental rules often exhibit drawbacks, including system specificity, limited robustness, and insufficient reproducibility \cite{hofer_engineering_2021}. Developing more versatile and efficient tools for biological design has the potential to extend the reach and impact of experimental techniques.

Inverse design tasks are usually translated into optimization problems, wherein the objective is to minimize the discrepancy between a property of the simulated system state and its desired value.
Pioneering research showcased the potential of using evolutionary algorithms to design genetic networks with specified functions \cite{francois2004}and and gaining insights into collective developmental cellular behaviors like segmentation \cite{francois2007}.
While sophisticated computational physical models of cell behavior do exist \cite{osborne2017, runser2024}, their applicability to reverse engineering remains limited, particularly in high-dimensional parameter spaces where parameter sweeps and evolutionary searches become inefficient. Automatic differentiation (AD) algorithms, on the other hand, enable efficient calculation of sensitivities of observables with respect to input parameters.  While originally developed for training large neural networks \cite{anil2023palm,brown2020language},
the same technical infrastructure can be used to differentiate through simulations of physical processes, offering gradient-based optimization as a viable alternative to tackle hard inverse problems of various scientific interest. Successful examples include the design of self-assembling materials \cite{goodrich_designing_2021}, the establishment of data-driven discretizations for partial differential equations \cite{bar-sinai_learning_2019}, nonequilibrium statistical mechanics \cite{engel2023optimal} and the design of quantum devices \cite{vargas-hernandez_fully_2021}. Neural Cellular Automata \cite{mordvintsev_growing_2020} represented an initial demonstration of how these methods could be used to engineer collective behaviors inspired by biology in lattice models. Hiscock, instead, showed that AD could be used to successfully learn meaningful couplings in ODE models of genetic networks \cite{hiscock2019}.

\begin{figure*}
\begin{center}
\includegraphics{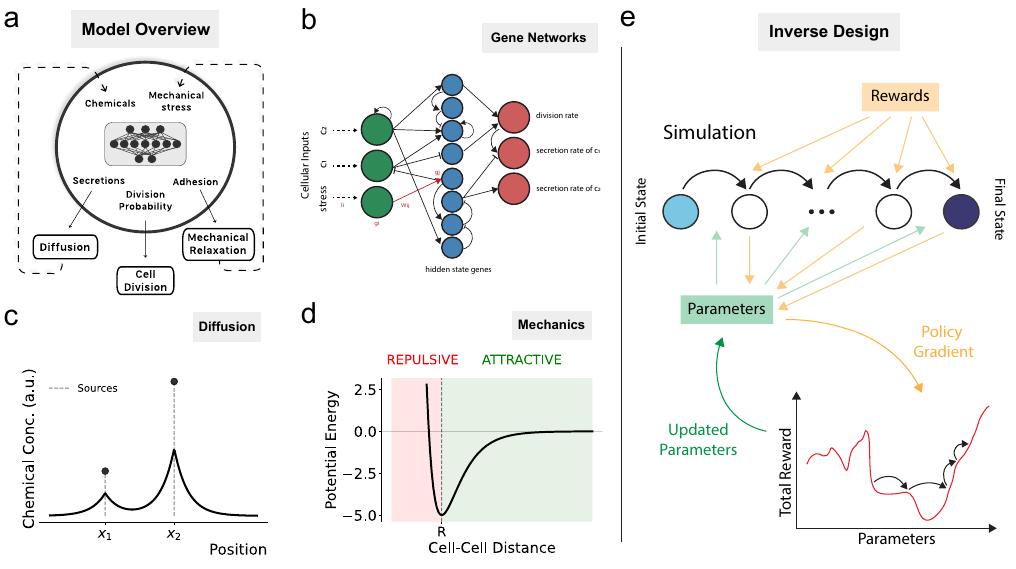}
\caption{\textbf{Modeling and Optimization Framework.} (a) Model overview. Decision-making circuits in cells sense local chemical and mechanical inputs to regulate secretion, division probability, and adhesion. The environment responds to cellular outputs by handling chemical diffusion, cell division, and mechanical relaxation. (b) Input signals and internal state are integrated and elaborated in each cell by internal gene regulatory networks. Green gene nodes sense inputs, blue genes are intermediates and red genes can modify cellular actions. (c) Released chemical factors diffuse in the environment following steady-state diffusion. (d) Mechanical interactions between cells are modeled with a pairwise Morse potential. The repulsive region accounts for volume exclusion forces and the attractive region models cell adhesion. (e) Gradient calculation and optimization. Following the REINFORCE approach, the log probability gradient of each time step is weighted by action rewards to estimate the average gradient. Parameters are updated through gradient ascent to maximize the expected reward.}
\label{fig:model}
\end{center}
\end{figure*}

In this study, we consider a tissue model composed of cells capable of division, growth, mechanical stress sensing, and morphogen excretion and detection. Cellular decisions are driven by responses to the local environment through an internal genetic network (Fig. \ref{fig:model}a).
We demonstrate that, starting from random or uninformative initializations, we can optimize the parameters governing genetic interactions to achieve non-trivial developmental scenarios in cell clusters. Our examples draw inspiration from various biological processes: axial elongation during limb bud outgrowth, maintenance of cell type homeostasis through chemical signaling in macrophage-fibroblast co-cultures, and growth homogenization via chemical signaling and mechanical stress response in the Drosophila wing disc.
Our framework enables the probing of the complex relationship between individual capabilities of proliferating cells and emergent properties of the resulting tissues.

\section*{Model and Optimization Framework}

\subsection*{Forward Model}
We model proliferating cells in a tissue with adhesive soft spheres, demonstrating examples of both two- and three-dimensional systems. While we present specific examples in this paper, our methods are potentially applicable to various physical models of cell growth and interaction. In our model, cells grow at a fixed rate until reaching a maximum radius, leading to continuous mechanical reorganization. Cells secrete morphogens, measure local concentrations of these chemicals, and sense mechanical stress. This locally sensed information guides decisions on chemical secretion modulation and division probability (Fig. \ref{fig:model}a). Cell division events are stochastic and produce two daughter cells, each with half the volume of a fully grown mother cell; the division plane is chosen uniformly at random. Decision-making circuits are modeled as regulatory networks comprising interacting genes with trainable excitatory or inhibitory couplings (Fig. \ref{fig:model}b). The regulatory network processes the information sensed by individual cells to output division propensity, rates of chemical secretion, etc. In most examples, the regulatory network weights are shared by all cells, though we also explore one scenario where two cell types have separate regulatory networks. Secreted chemical factors allow information sharing by diffusing in the environment (Fig. \ref{fig:model}c). Cells mechanically interact through a Morse potential (Fig. \ref{fig:model}d), with a repulsive part for  excluded volume interactions and an attractive part for cell-cell adhesion \cite{koyama_effective_2023}. Simulations are conducted for a specific number of cell divisions, so the final state cannot in general be considered a steady state. Further mathematical detail can be found in the \textit{Materials and Methods} section and in the \textit{Supplementary Information}.

\subsection*{Inverse Design}
Our objective is to discover local growth rules that enable cells to develop into a final configuration with desired properties. To effectively navigate the high-dimensional, non-convex parameter space of local interaction rules, we employ gradient descent schemes, widely used in modern machine learning for such optimizations. Our entire simulation is written to be automatically differentiable, leveraging recent advances from the deep learning and reinforcement learning communities. We utilize the JAX library \cite{jax2018github} for efficient numerical and automatic differentiation algorithms, and other JAX-based libraries such as Equinox \cite{kidger2021equinox} and JAX-MD \cite{jaxmd2020} for molecular dynamics. 

We employ score-based methods such as REINFORCE \cite{Sutton1998} to address the intrinsic stochasticity of proliferation dynamics, which implies that the loss function is not directly differentiable (Fig. \ref{fig:model}e). In the REINFORCE update, rewards and penalties are assigned to each action taken (i.e. each cell division) to provide the weighting signal for the gradient of the trajectory probability. This weighted gradient is then used to nudge the weights of the gene network in the right direction. In practice, the system gradually learns which division events are the most favorable and increases their probability for the next simulation. To guide the reader possibly unfamiliar with these concepts, we also include a set of in-depth tutorial notebooks in the GitHub repository \cite{jax_morph_tutorials} and a longer theoretical explanation in the \textit{Supplementary Information}. Since the dynamics of the environment, including chemical secretion and diffusion, depend partly on the same parameters (the gene network couplings) that cells use to decide whether to divide or not (their “policy”), it is essential to evaluate the gradient of the environment dynamics with respect to the gene network parameters. This necessity makes automatic differentiation particularly crucial in this context. Once the necessary gradients are calculated, optimization is performed using well-established gradient descent schemes, such as Adam \cite{kingma_adam_2017}. The learned networks are then further simplified by removing all edges with weights smaller than an experiment-dependent threshold, such that the functional backbone of pruned network be highlighted while preserving the system properties we optimized for.

\begin{figure*}
\centering
\includegraphics{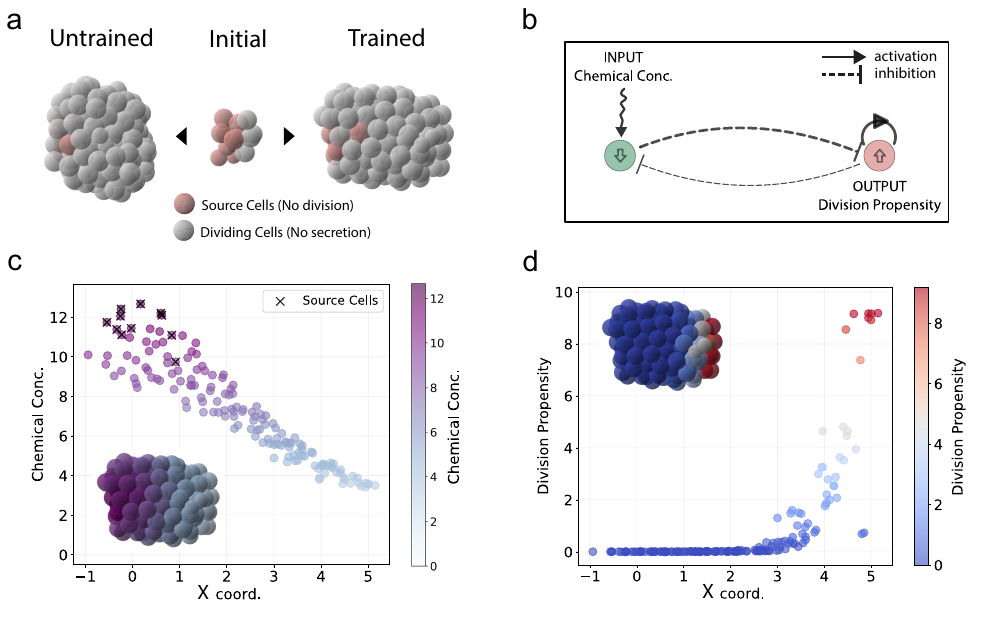}
\caption{\textbf{Elongation.} Maximization of the squared sum of the x-coordinates of cells, resulting in horizontal elongation in the optimized cluster. (a) On the left, the final configuration of a simulation with randomly initialized parameters; on the right, the final simulation state after learning. Source cells (in red) secrete the growth factor and cannot divide. Proliferating cells (in gray) sense the growth factor and divide in response to it. (b) The learned gene network. The receptor gene is activated only by the presence of the external chemical factor, which results in repression of the division propensity. Edge boldness is correlated to connection strength. (c) Chemical gradient created by source cells along the cluster x-coordinate. In the inset, the same gradient visualized in physical space. (d) Division propensity distribution at the end of the simulation along the x-axis, highlighting the concentration of dividing cells at the tip.}
\label{fig:elongation}
\end{figure*}

\section*{Spatial Control of Tissue Growth}

Understanding how to program the growth of complex shapes can aid in our comprehension of developmental body plans driven by chemical patterning. Rationally designed synthetic gene circuits have been shown to be able to drive the formation of target multicellular structures found in morphogenesis \cite{toda_engineering_2020, guo_programming_2021}. We investigate here how to use our framework to discover decision-making circuits that utilize local chemical interactions to grow target shapes. 
As an example, we focus in this section on symmetry-breaking axial elongation. This process is the first geometric-defining process to happen in many developmental programs and is vital for establishing, for instance, the antero-posterior axis in animals \cite{darras2018}. We learn a gene circuit that enables cells to chemically interact with a non-proliferating population of "source cells" to elongate along a given axis. Similar problems have recently been explored by engineering, for example, contact-based cell communication instead of chemical communication \cite{lam2022, courte2024}.
We consider a three-dimensional model of dividing cells, that we optimize to minimize the squared sum of the x-coordinates of cells, thus requiring the cluster to elongate horizontally (Fig. \ref{fig:elongation}a). The system is composed of two different cell types. One secretes a diffusible factor (source cells, in red in Fig. \ref{fig:elongation}a), but cannot proliferate. The other cell type (in gray in Fig. \ref{fig:elongation}a) can only sense the factor concentration and modulate its division propensity accordingly.  Local chemical concentration is the input to the proliferating cells' gene network, allowing them to infer positional information through local physical cues.

The learned mechanism (Fig. \ref{fig:elongation}b) drives a dynamic process reminiscent of limb bud outgrowth in animal development. Secretion of the chemical factor by source cells results in a steady-state chemical gradient along the cell cluster (Fig. \ref{fig:elongation}c). Due to the network's input-to-output strong inhibitory link, regions of higher chemical concentration, proximal to the source cells, experience  a decrease in the division propensity of proliferating cells. Conversely, areas with lower chemical concentration experience less inhibition, leading to a sustained higher division propensity located in the distal region. In the initial spherical cluster, source cells are localized on the left and proliferating cells on the right, the latter cells possessing a non-zero initial division propensity. Proliferating cells closer to the source will experience a diminishing division propensity over time due to the presence of the chemical factor. As these cells divide, they preferentially expand the cluster in directions of lower chemical concentration, effectively moving away from the source and further reducing the chemical concentration at the extending tip. This creates a self-reinforcing loop: cell division events are concentrated at the distal tip where the chemical signal is weakest, ensuring directional growth and elongation away from the source of the secreted factor (Fig. \ref{fig:elongation}d). 
Crucially, the weak inhibitory feedback from division propensity back to chemical input acts as a contrast enhancer, sharpening the division response to the chemical gradient and producing a ''winner-takes-all`` kind of behavior. In areas of high chemical concentration, the strong inhibitory forward connection suppresses division, thus diminishing the input suppression by the weaker feedback connection and leading to even stronger division suppression. Conversely, in regions of low chemical concentration where direct inhibition is less prominent, division propensity increases and the weak feedback becomes functionally significant, further reducing the perceived chemical input. This effect amplifies the division propensity output specifically in these low-concentration regions. The positive self-loop on division propensity further stabilizes and amplify this directional growth by sustaining high division rates in the low-concentration tip regions, and working in synergy with the contrast-enhancing feedback to create a robust and spatially well-defined elongation.

While significantly simplified compared to biochemical networks governing directional growth, the learned architecture functionally recapitulates the core logic of some developmental processes. Like limb bud outgrowth, it achieves directional elongation through localized growth at the tip \cite{boehm2010}. Unlike the growth-promoting signaling factors (like FGF) secreted by the Apical Ectodermal Ridge, though, it employs an inhibitory regulatory signal originating from the base to restrict proximal proliferation and promote growth at the distal end. 
This regulation is learned consistently in all our optimization runs. The lack of alternative mechanisms (e.g. maintaining a proliferating zone close to the source cells) might be due to different reasons, like the particular mechanical properties chosen to simulate the system, the absence of external geometric constraints or the specific reward function chosen.

\begin{figure*}
\centering
\includegraphics{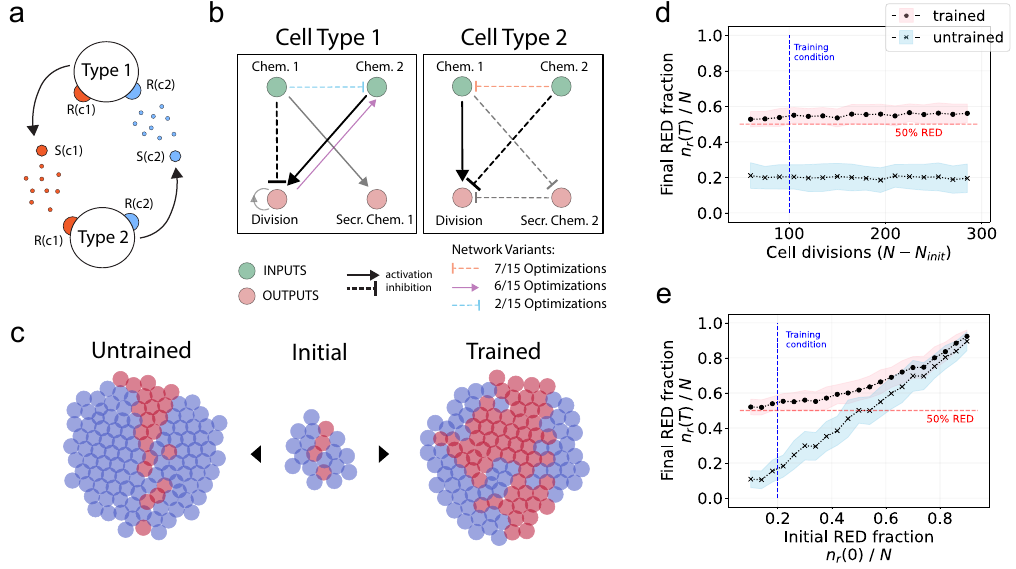}
\caption{\textbf{Chemical Regulation of Homeostasis.} (a) Two cell types each exclusively secrete a chemical factor that can be sensed by both cell types. (b) Visualization of the learned gene networks. Interactions in black are present in every learned network. Colored interactions are single variants that are present only in a subset of learned solutions, with the reported frequency. This gives three type 1 network variants and two type 2 variants, totaling to 6 possible overall regulatory combinations. Edge boldness is correlated to connection strength. (c) An initially imbalanced state (in the middle) grows into an equally unbalanced final state with untrained network parameters (left) and into a balanced state with learned parameters (right) (d) Average over 100 realizations of final proportion of red cells at the end of simulations of increasing lengths. (e) Average over 100 realizations of final proportion of red cells at the end of simulations with increasing proportions of red cells in the starting state. In both plots the shaded area corresponds to average plus or minus one standard deviation.}
\label{fig:homeostasis}
\end{figure*}

\section*{Chemical Regulation of Tissue Homeostasis}

Maintaining stable ratios of different cell populations is crucial for tissue function, and disruptions of this balance can lead to disease \cite{yap2021}. To maintain these stable proportions, cellular circuits must prevent uncontrolled growth of any single cell type. Cell communication mediated by  diffusible factors constitutes a key mechanism for achieving tissue homeostasis and regulating tissue composition.

One prominent example is the interplay between osteoblasts and osteoclasts in bone homeostasis \cite{kim2020}. Bone-forming osteoblasts primarily secrete RANKL, which promotes osteoclast differentiation and activation. Conversely, osteoclasts secrete their own factors (like S1P) and release bone matrix factors (e.g. TGF-$\beta$) during bone resorption, influencing osteoblast abundance. This reciprocal signaling establishes a feedback loop that maintains the balance between bone formation and resorption. 
Similar phenomena are also observed \textit{in vitro}. For instance, co-cultures of macrophages and fibroblasts can spontaneously reach stable ratios, even from highly imbalanced starting conditions \cite{zhou_circuit_2018}. Both cell types secrete growth factors that selectively stimulate the growth of the other: macrophages primarily produce PDGF and fibroblasts secrete Csf1, while other diffusible factors contribute to self-regulation of growth. Cell-to-cell chemical communication via diffusible factors has also been shown to be crucial for regulating homeostasis during cell differentiation \cite{saiz2020, raina2021}.

Inspired by the macrophage-fibroblast system in particular, we investigated chemical mechanisms capable of maintaining a balanced 50/50 ratio between two cell types in a growing cluster. In this 2D model, two cell types interact chemically, each secreting a distinct chemical factor (Fig. \ref{fig:homeostasis}a). Cell behavior is governed by cell-type-specific genetic networks, learned through optimization, which regulate division propensity and secretion of diffusible factors (Fig. \ref{fig:homeostasis}b). As shown in Fig \ref{fig:homeostasis}c (center),  simulations begin with an imbalanced initial state, comprising $n_r=4$ red cells (cell type 1) and $n_b=16$ blue cells (cell type 2). The untrained network is initialized to yield uniform division propensities: after 100 cell divisions, the initial imbalance is found again in the final state (Fig \ref{fig:homeostasis}c, left). The learned mechanisms, instead, successfully yield balanced cell type populations at the end of the simulation. Gene couplings are optimized to minimize the number difference between the two cell types during growth (Fig \ref{fig:homeostasis}c, right).

$$ \mathcal{L}_t = |n_{r}(t) - n_{b}(t)| $$
is the loss at time $t$, where $n_r$ and $n_b$ are the number of red and blue cells respectively, and $n_r + n_b = N$ is the total cell number.

The learned networks for each cell type are shown in Fig. \ref{fig:homeostasis}b.  Both cell types sense both chemicals, but each modulates the secretion of its own factor only. We focus first on the connections in black in the figure, representing the core "backbone" network consistently learned across optimization runs and sufficient for homeostasis. As expected, the strongest effects (bold in Figure \ref{fig:homeostasis}b) link input chemicals to proliferation regulation, in a mirrored manner across cell types. In both cell types, chemical secreted by the same type inhibits division, while chemical secreted from the other type promotes it. Secretion control in both cell types is governed solely by chemical factor 1, associated with the initially less abundant red cell type. In cell type 1, chemical 1 simply enhances its own production. In cell type 2, secretion regulation is less straightforward: chemical 1 inhibits secretion of chemical 2, and high secretion of chemical 2 contribute to proliferation suppression.

Consider a region where type 2 cells are surrounded by their own type, levels of chemical 1 are low and chemical 2 is secreted at a basal rate. In this case, the lack of chemical 1 removes suppression of chemical 2 secretion, resulting in higher chemical 2 levels. High levels of chemical 2  and strong secretion activate division inhibitory pathways for type 2, halting proliferation. Conversely, when sensed levels of chemical 1 rise due to increased division of cell type 1, the inhibition of chemical 2 secretion is lifted, eliminating in turn the secondary inhibitory feedback on proliferation. Simultaneously, the strong positive regulatory connection between chemical 1 and division promotes type 2 proliferation. Increased type 2 division and secretion of chemical 2 again activates the inhibitory division pathways, and the cycle repeats. Notably, spatial structure plays a key role: cell divisions place daughter cells near parent cells, altering neighborhood cell type composition. This increases the local concentration of the self-secreted chemical and, at the same time, decreases the concentration of the other chemical as cells move away from surrounding sources. Despite model differences, the learned solutions share key features with the macrophage-fibroblast homeostasis circuit reported in \cite{zhou_circuit_2018}. Both systems feature mutual proliferation promotion between the two cell types and negative growth self-regulation. Furthermore, also in this case one of the two factors negatively regulate the other.

Beyond the most common networks, optimization also identifies two alternatives for cell type 1 and one for cell type 2. These variants differ from the most frequent architecture by single additional edges (colored differently in Figure \ref{fig:homeostasis}b). Network variants are not correlated between cell types and can appear in any of the six possible combinations. The weaker additional links enhance existing dynamical behaviors without fundamentally altering network function. Their presence may reflect optimization convergence to local minima or variations in convergence speed, rather than specific functional advantages. 

Remarkably, the learned gene regulations exhibit robustness and generalization beyond the explicit optimization goals. If one fixes the initial imbalance to the one used during training ($n_r/N = 0.2$), the learned regulation maintains homeostasis even in extended simulations far beyond the training duration of 100 divisions, showing no degradation (Fig. \ref{fig:homeostasis}d). Furthermore, maintaining 100 cell divisions but varying the initial proportions of cell types, homeostasis is reasonably maintained up to nearly twice the initial imbalance (Fig. \ref{fig:homeostasis}e). Beyond this point, the regulatory circuit starts to overcompensate and loses its advantage over random cell divisions. This behavior aligns with the regulatory asymmetry observed in the discovered networks, reflecting the fact that all initial simulation states during optimization are imbalanced in the same direction (i.e. with more blue cells than red ones). We suppose that the initial tissue state results from an upstream cell differentiation process which does not produce completely random cell type proportions, so having no initial variability seems like a reasonable approximating assumption.

\begin{figure*}
\begin{center}
\includegraphics{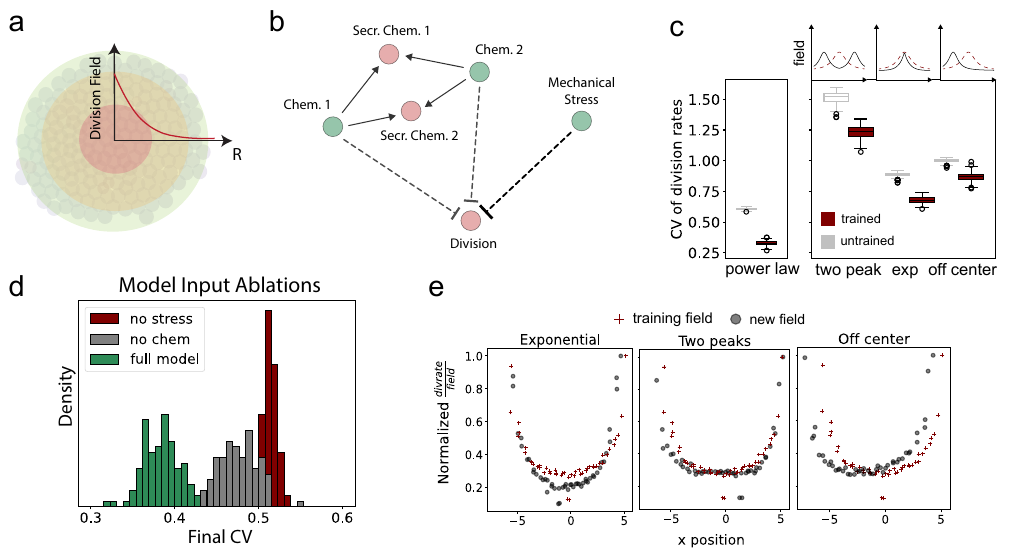}
\caption{\textbf{Mechano-chemical Regulation of Homogeneous Growth.} (a) Scheme of the problem setup, inspired by wing disc growth in \textit{Drosophila}. The extrinsic division scaling mirrors the experimentally observed tendency of cells to divide in the center of the cluster. (b) The learned gene network. Division is controlled both by the external division field (left implicit in the diagram) and by the other cell inputs. Edge boldness is correlated to connection strength. (c) Proliferation regulation learned in a power-law division field yield lower coefficient of variation of division propensities also in systems with different proliferation biases. Left to right: the power-law field used in training, a field with two peaks instead of one, an exponential field, and a field with an off-center peak. Distributions are calculated over 100 simulations. (d) Ablation studies assess the importance of both mechanical and chemical regulation. Separately removing chemical or mechanical feedback leads to more proliferation heterogeneity with respect to the presence of both.(e) Division rate divided by the extrinsic division field is plotted for each kind of simulation in panel (c). This shows that cells regulate division probabilities by adapting to the shape of the extrinsic division field.}
\label{fig:homog_growth}
\end{center}
\end{figure*}

\section*{Mechanical Control of Cell Proliferation}

Mechanical signaling plays a crucial role in morphogenesis, influencing both individual cells and tissue-level processes. At the cellular level, mechanosensation affects gene expression through various biochemical pathways that transduce forces \cite{agarwal_mechanosensing_2021}. On a broader scale, mechanical signals convey local instructions to shape global structures \cite{agarwal_mechanosensing_2021, irvine_mechanical_2017, trinh_how_2021, legoff_global_2013}. In tissues, mechanical interactions can operate over larger distances than biochemical signals due to tension propagation through connected cells \cite{legoff_global_2013}. Furthermore, the accumulation of local mechanical stress can impact the global growth of developing structures. For instance, mechanical feedback has been proposed as a mechanism for growth arrest in the Drosophila wing imaginal disc \cite{shraiman_mechanical_2005, hufnagel_mechanism_2007}. In this study, we explore these concepts by optimizing a model inspired by the Drosophila wing disc development to uncover mechanisms that couple mechanics to growth.

Our model incorporates a constant radial growth factor gradient that modulates division propensity, with cells closer to the cluster center having a higher probability of division (Fig. \ref{fig:homog_growth}a). This aligns with experimental observations of transiently faster growth rates in central cells during early wing disc development \cite{tripathi2022}. This extrinsic  division scaling, fixed throughout the simulation, follows a power-law decay. The total division propensity of cell $i$ is given by:

$$
d_i = g_i(stress, chemicals)*f(r)
$$

where $g_i(stress, chemicals)$ is the division output of the internal gene network and $f(r) \propto 1/r^2$ is the radial scaling due to the external factor. 

As the cell cluster expands, each cell modulates its division propensity through their internal gene circuit, receiving sensed chemicals and stress as inputs. We aim to identify local growth rules that achieve homogeneous proliferation throughout the cluster despite the nonuniform effect of the external factor gradient. This optimization for uniform growth rates is inspired by experimental evidence, showing similar growth rates across different regions of the wing disc during most of its development, despite small-scale local variations \cite{tripathi2022}.

We optimize couplings between genes in the circuit to minimize the coefficient of variation of division propensities:

$$
\mathcal{L} = \text{CV}[\textbf{d}(T)] = \frac{\sigma [\textbf{d}(T)]}{\mu [\textbf{d}(T)]}
$$
where $\textbf{d}(T)$ are the division propensities at final time $T$.

The optimized gene circuit utilizes both chemical and mechanical information to modulate division propensity (Fig. \ref{fig:homog_growth}b). This regulation leads to a significantly lower CV of cells' division rates (Fig. \ref{fig:homog_growth}c). Remarkably, the gene circuit optimized on the power law gradient successfully reduces growth variations also when exposed to external division scaling fields unseen during training. To assess the robustness of the mechanism, we expose the trained gene circuit to different external gradients: an exponentially decaying gradient, a gradient with two symmetric peaks, and a gradient with a single asymmetric peak.(Fig. \ref{fig:homog_growth}c, left panel).

While simplified, our model parallels several phenomena observed in experimental studies of wing disc growth regulation and captures the essence of mechanical feedback in growth regulation. In the learned network, mechanical stress directly inhibits cell division, mirroring the role of the Hippo pathway in the wing disc. Experimentally, mechanical stress has been shown to activate the Hippo pathway \cite{tripathi2022}, leading to Warts-mediated phosphorylation and inactivation of the growth-promoting transcription co-activator Yorkie (Yki), ultimately suppressing cell proliferation. Conversely, reduced mechanical stress allows Yki to remain unphosphorylated, enter the nucleus, and drive cell proliferation necessary for proper tissue growth and development. Our model represents a coarse-grained version of this mechanism through the inhibition of division propensity by the stress-sensing gene.

Although the secreted factors in our model lack direct biological counterparts, they serve as a simplified abstraction of the complex chemical signaling networks coordinating growth control in the Drosophila wing imaginal disc. The incorporation of secreted factors for feedback regulation conceptually represents the broader principle of growth control via morphogen gradients in the wing disc. For instance, the Decapentaplegic (Dpp) and Wingless (Wg) proteins form concentration gradients across the anterior-posterior and dorsal-ventral axes respectively, regulating growth and patterning \cite{tripathi2022}. While not direct analogues of these specific morphogens, our model's diffusible factors capture the essence of spatially distributed chemical signals that influence growth patterns. The importance of these factors is further demonstrated in Fig. \ref{fig:homog_growth}d, where we compare the performance of circuits with ablated inputs. Removing the chemical factors significantly impairs the cluster's self-regulation capabilities, confirming that both chemical and mechanical regulation are essential in this context. Moreover, the gene network outputs appear to track both the shape of the fields and patterns of stress accumulation (Fig. \ref{fig:homog_growth}e, Supp Fig. S6c), consistent with a mechanical feedback mechanism.

\section*{Additional Experiments}
We report in the \textit{Supplementary Information} a set of additional experimental setups that we omitted from the main text to focus on the most biologically relevant findings. We provide here a brief summary of such examples, to further stress the flexibility of our method and possibly inspire further applications.
When cells have access to estimates of chemical gradients constructed by communicating with nearest neighbors, they can learn to elongate in one direction (Fig. S2) and grow in a branched shape (Fig. S3) by only including a single cell type. In these cases they learn to create emergent chemical gradients without the need for external asymmetric sources and show remarkable robustness to cell removal. 
We also show that, besides chemical secretion and division propensity, we can endow cells with control over their adhesion strength. By regulating a set of homotypic and heterotypic ``cadherins", cells can learn basic assembly programs for checkerboard, lobed and core-shell structures (Fig S1).

\section*{Discussion}
Our results demonstrate the power of gradient-based optimization methods in discovering mechanisms within the space of physical models for collective cell behavior. By taking gradients through a physical model of interacting cells written using automatically differentiable code, we were able to learn decision-making programs within individual cells to orchestrate growth into target shapes or with other target properties. We were able to overcome challenges associated with differentiating through the discrete nature of cell division events and non-differentiable loss functions with an approach based on classic reinforcement learning. The mechanisms that were learned took the form of gene regulatory networks, which readily offer a physical and biological interpretation. The learned mechanisms were local in nature: cells did not have access to global information (e.g. their absolute position) and had to infer it instead by sensing chemicals and mechanics and communicating with nearby cells through these physical channels. \\
The gene networks optimized here could be substituted with more detailed biophysical models or constrained to incorporate prior biological knowledge. Furthermore, the soft sphere model we have presented here is simply one of the possible choices. The physical details can be modified to optimize over an alternative bespoke forward model with the same methods we demonstrated in this work. Questions about minimal mechanisms, robustness to variation and perturbations, number of unique solutions or growth principles could also be framed and investigated within the same setting. 
\\
Our results are a promising first step towards inferring mechanisms in existing multicellular systems or learning how to program them for desired behaviors. In particular, morphogenetic mechanisms frequently utilize local feedback to coordinate intricate forms, often through chemical or mechanical signals. Examples include a myriad of developmental systems, from the development of the wing imaginal disc in Drosophila to the periodic segmentation of blocks in somitogenesis. The development of robust and useful organoid systems also requires an understanding of spatiotemporal chemical and physical cues needed to guide the growth of stem cells. Physical models could be optimized to reproduce experimental data from time-lapse imaging of such developmental systems to gain insights into individual decision making functions that recapitulate complex behaviors. Gene expression datasets could be incorporated to bridge the gap between known and unknown information in biological systems. Maximization or minimization principles could be explored beyond simplified models. Our work opens up a multitude of avenues for further explorations of the landscape of growth programs for desired emergent behaviors in living systems. \\

\section*{Materials and Methods}

Fully differentiable, hardware-accelerated simulations were coded using the JAX python library \cite{jax2018github} and other libraries in the JAX ecosystem. In particular, JAX-MD \cite{jaxmd2020} was used to perform molecular dynamics simulations for cell mechanics and Equinox \cite{kidger2021equinox} as the simulation backbone.

We proceed to give a very succinct overview of the main components of our model. A full mathematical account of simulation and optimization details is given in the \textit{Supplementary Information}.

\subsection*{Cell division and growth} Each cell $i$ is endowed with a \textit{division propensity} $d_i$ calculated by the internal gene network. One cell divides at each time step, with probability:

$$ P(i\ \text{divides}) = \frac{d_i}{\sum_{i} d_i} $$
The division substitutes one mature cell with two newborn cells with half the volume. The angular orientation of the division plan is chosen uniformly at random. 

Cells' radii $R_i$ grow linearly in time up to a maximum radius $R_{max}$:

$$
R_i(t+\Delta t) = \min(R_i(t) + \Delta R, R_{\text{max}})
$$

\subsection*{Mechanical interactions} Cells are modeled as adhesive soft spheres. Interactions are defined by the Morse energy potential:

$$
V_{ij}(r) = \epsilon_{ij} \left[ 1 - e^{-\alpha_{ij}(r_{ij} - \sigma_{ij})} \right]^2 - \epsilon_{ij}
$$
where $\epsilon_{ij}$ is the well depth for the potential between particle $i$ and $j$, $\alpha_{ij}$ is the range of interaction, $\sigma_{ij}$ is the sum of radii of particles $i$ and $j$, and $r_{ij}$ is the distance between the two particles.

\subsection*{Diffusion} We assume that chemicals diffuse freely in the environment and are taken up by cells at a constant rate. It is also assumed that diffusion is much faster than all other processes and can be therefore be considered at steady state. For any one chemical $k$:

$$
\frac{\partial c_k}{\partial t} = D_k \nabla^2 c_k - K_k c_k + S_k = 0
$$
The equation is discretized and solved on the lattice formed by the cells' centers with the graph Laplacian technique.

\subsection*{Genetic regulatory interactions} We model internal cell decision functions with a simple ODE model of gene regulation inspired by Hiscock \cite{hiscock2019}:

$$
\frac{dg_i}{dt} = \phi(\Sigma_j W_{ij}g_j + b_i) + I_i - k_i g_i
$$
Considering the $i$-th gene, $g_i$ is the expression level, $W_{ij}$ is the regulatory influence of gene $j$ on $i$, $b_i$ determines the gene transcription rate in absence of regulation, $k_i$ is the degradation rate, and $I_i$ is a forcing signal coming from one of the sensed cell inputs. $\phi$ is a sigmoid function that ensures positive transcription rates saturating at high input levels.

\subsection*{Gradient calculation} 
Due to the presence of discrete stochastic events, the gradients of simulations involving cell division events cannot be calculated by direct application of the chain rule. Following REINFORCE \cite{Sutton1998}, we can estimate the gradient of the expected loss for a single simulation:

$$
\nabla_\theta \mathcal{L}(\theta) \approx \sum_{t=0}^{T-1} L_t \nabla_\theta \log \pi_\theta(a_t|s_t)
$$
where $L_t$ is the discounted future loss, $\theta$ represents the trainable gene network parameters and $\pi$ is the probability of the division event parametrized by $\theta$.

\acknowledgements{This work is dedicated to the memory of Alma Dal Co. We thank for insightful discussions: the members of the Brenner Group, the members of the former Dal Co group, Boris Shraiman and Dillon Cislo. R.D. thanks Herv\'{e} Turlier for helpful discussions. This work was supported by  the Office of Naval Research through grant numbers ONR N00014-17-1-3029, ONR N00014-23-1-2654 and the NSF AI Institute of Dynamic Systems
(2112085).}

\section*{Data Availability}
The custom library written to conduct this study is available at \href{https://github.com/fmottes/jax-morph/}{https://github.com/fmottes/jax-morph/}, along with tutorial notebooks.

\bibliography{refs}

\end{document}


\title{SUPPLEMENTARY INFORMATION\\Engineering morphogenesis of cell clusters with differentiable programming}

\author{Ramya Deshpande$^*$, Francesco Mottes$^*$, Ariana-Dalia Vlad, Michael Brenner and Alma Dal Co}

\maketitle

\tableofcontents

\newpage

\section{Forward Simulation}
We run simulations using JAX-MD \cite{jaxmd2020}, a molecular dynamics engine built using the automatic differentiation framework JAX \cite{jax2018github}. Cells are particles instantiated in "free" space, with a set of properties represented by a \texttt{CellState} datatype; this object contains the position of particles, their radii, division propensities, chemical concentrations at the center of each cell and any additional state information appropriate for a simulation (for example: stress, finite difference gradients of the chemical concentrations, etc.). At the start of the simulation, arrays of size (T, n + T, Dim) are instantiated for each property in \texttt{CellState}, where $T$ is the number of timesteps, $n$ is the initial number of cells, and $Dim$ is the dimension of the cell property (2D or 3D). Each simulation timestep consists of one cell division, so the final configuration will contain $N = \text{n} + \text{T}$ cells.  \\
To create an initial state with $n$ cells, we place a cell at the origin, and run $n - 1$ iterations of cell division, cell growth, and mechanical relaxation to generate a cluster of the desired initial size. All division propensities are equal in the initial state, so each cell is chosen to divide with equal probability, generating a radially symmetric cluster. 
The actual simulation can be customized to contain any subset or combination of the steps detailed below; the specifics of the simulation used for each instantiation in the main text is detailed in the \textit{Experimental Details} section. The code used for the simulations can be found in the following GitHub repository: \href{https://github.com/fmottes/jax-morph}{https://github.com/fmottes/jax-morph}.

\subsection*{Cell Growth} All cell radii grow constantly in time, up to a specified maximum radius $R_{\text{max}}$:

$$R_i(t+\Delta t) = \min(R_i(t) + \Delta R, R_{\text{max}})$$

This operation can also be made differentiable if necessary, by means of a suitable smoothing function in place of the hard \textit{min} operation.

\subsection*{Mechanical Interactions} 
Cells mechanically interact with one another via a pairwise Morse potential. The potential has a minimum at a distance defined by the sum of the two interacting cells' radii. It is very repulsive at shorter distances (\textit{volume-exclusion interactions}) and slightly attractive at longer distances (\textit{cell-cell adhesion}). The range of attraction, the strength of the exclusion and the stiffness of the interaction can be modified by changing the potential parameters.

Relaxation is performed by gradient descent energy minimization of the Morse potential for a fixed number of steps (except in the case where we learn cell adhesion). The pairwise potential is defined by the following expression:

$$
V_{ij}(r) = \epsilon_{ij} \left[ 1 - e^{-\alpha_{ij}(r_{ij} - \sigma_{ij})} \right]^2 - \epsilon_{ij}
$$
where $\epsilon_{ij}$ is the well depth for the potential between particle $i$ and $j$, $\alpha_{ij}$ is the range of interaction, $\sigma_{ij}$ is the sum of radii of particles $i$ and $j$, and $r_{ij}$ is the distance between the two particles. In most simulations, we assume for simplicity that $\alpha_{ij} = \alpha$ and $\epsilon_{ij} = \epsilon$. This need not be the case in general. In particular, in the example where we learn the regulation of cell-to-cell adhesion, the values of $\epsilon_{ij}$ are determined by the cells' internal regulatory mechanisms.

\subsection*{Cell Division}
Each cell $i$ is endowed with a \textit{division propensity} $d_i$ calculated by the internal gene network and stored in the \texttt{CellState}. One cell divides at each time step, with probability:

$$
P(i\ \text{divides}) = \frac{d_i}{\sum_{j} d_j}
$$
A cell $i$ is first randomly chosen to divide with these probabilities. A random direction is then selected as $\theta \in [0, 2\pi]$. The positions of the daughter cells are calculated as:

\begin{gather}
\Bar{r}_{1,x} = \Bar{r}_{0,x} + R_{\text{birth}}\text{cos}(\theta), \, \Bar{r}_{1,y} = \Bar{r}_{0,y} + R_{\text{birth}}\text{sin}(\theta) \\
\Bar{r}_{2,x} = \Bar{r}_{0,x} -R_{\text{birth}}\text{cos}(\theta), \, \Bar{r}_{2,y} = \Bar{r}_{0,y} - R_{\text{birth}}\text{sin}(\theta)
\end{gather}
where $(\Bar{r}_{1,x}, \Bar{r}_{1,y})$ and $(\Bar{r}_{2,x}, \Bar{r}_{2,y})$ are the positions of the daughter cells, $(\Bar{r}_{0,x}, \Bar{r}_{0,y})$ is the position of the mother cell and $R_{\text{birth}}$ is fixed as the radius of newly born cells.  Each daughter cell inherits all other properties of its mother cell.

\subsection*{Diffusion}
We assume that chemicals diffuse freely in the environment and are taken up by cells at a constant rate. It is also assumed that diffusion is much faster than all other processes and can be therefore be considered at steady state. Concentrations of a chemical $k$ (in arbitrary units) at each cell site are determined by the following equation:

$$
\frac{\partial c_k}{\partial t} = D_k \nabla^2 c_k - K_k c_k + S_k = 0
$$

The equation is discretized and solved on the lattice formed by the cells' centers with the graph Laplacian technique. The PDE above is first reformulated by substituting the diffusion operator with the discrete graph Laplacian $L$. This results in a linear system of ODEs for each chemical, describing the concentration of a chemical on each lattice site. The equation is then solved for the steady state concentrations $\{\mathbf{c}_k\}$:

$$
\frac{\partial \mathbf{c}_k}{\partial t} = D_k L \mathbf{c}_k - K_k \mathbf{c}_k + \mathbf{S}_k = 0
$$   
$D_k$ is the diffusion coefficient of chemical $k$ and $K_k$ is the uptake rate for chemical $k$, the same for every cell for simplicity. $\mathbf{S}_k$ is an $N$ x $1$ vector of secretion rates of chemical $k$ by every cell, calculated by the gene networks. $\mathbf{c}_k$ is an $N$ x $1$ vector containing the concentration of chemical $k$ at all cell sites. The discrete graph Laplacian operator is constructed as:

$$   
L = \text{deg}(A) - A
$$
where $A$ is the adjacency matrix that describes the connections between the discrete spatial sites. $\text{deg}(A)$ is a matrix that has the sum of $A$'s rows (that is, the node degrees) as diagonal elements and is zero everywhere else. For all cells $i$:

$$
\{\text{deg}(A)\}_{ii} = \sum_{j=0}^N A_{ij} 
$$

Apart from the branching and homogeneous growth examples, we construct the graph Laplacian to simulate diffusion in a closed system with reflecting boundary conditions. In this case the adjacency matrix $A$ is calculated as the inverse of the pairwise distance between every pair of cells:

$$
A_{ij} = \frac{1}{\text{dist}(i,j)} = \frac{1}{\lVert \Bar{r}_i - \Bar{r}_j \rVert}
$$

In order to simulate permeable boundaries without the additional burden of having a higher resolution lattice only for chemical diffusion, we adopt an approximate approach. We now construct the adjacency matrix by connecting cells that are nearest neighbors with an edge of weight 1 (that is approximately their distance) and zero otherwise:

$$
A_{ij} = 
\begin{cases}
    1 & \text{if}\ \text{dist}(i,j) \leq R_i+R_j \\
    0, & \text{otherwise}
\end{cases}
$$
We then heuristically detect the nodes that are on the boundaries of the cluster by looking at the number of nodes they are connected to. We connect all of these boundary nodes to a ghost sink node that receives and dissipates chemicals, to simulate the effect of chemicals lost in the medium surrounding the cluster. In practice, this is done by adding, in the degree matrix, an extra unit of degree to the nodes on the boundary.

\subsection*{Chemical Gradients} Multicellular collectives can infer weak concentration gradients across the collective through individual cells measuring differences in chemical concentrations across their diameters \cite{endres_accuracy_2008}.
To achieve such an estimate of the spatial gradient of chemical $k$ sensed by cell $j$, a unit vector pointing to each nearest neighbor $i$ of cell $j$ is scaled by the concentration of chemical $k$ in those cells, and the the contributions from  all neighbors are summed up:

\begin{gather}
    \nabla c_k(\Bar{r}_j) \approx \sum_{i \in nn(j)} c_k(\Bar{r}_i)\frac{\Bar{r}_i - \Bar{r}_j}{\lVert \Bar{r}_i - \Bar{r}_j \rVert} 
\end{gather}
Here, $\Bar{r}_i$ is the position of cell $i$, and $c_k(\Bar{r}_i)$ is the concentration of chemical $k$ at position $\Bar{r}_i$.

\subsection*{Mechanical Stress}
As a proxy for the mechanical stress exerted by cell $i$ on cell $j$ in our 2D environment, the component-wise force on $i$ by $j$$, (F_{ij, x}, F_{ij, y})$, is multiplied by the component-wise unit vector pointing from $j$ to $i$$, \, (\Bar{r}_{j,x} - \Bar{r}_{i,x}, \Bar{r}_{j,y} - \Bar{r}_{i,y})$; this is summed over the components. For cell $i$ we compute:

$$
\sigma_i = 
    \Sigma_j \left[ F_{ij,x}\cdot \frac{\Bar{r}_{j,x} - \Bar{r}_{i,x}}{| \Bar{r}_{j,x} - \Bar{r}_{i,x} |} + F_{ij,y}\cdot \frac{\Bar{r}_{j,y} - \Bar{r}_{i,y}}{|\Bar{r}_{j,y} - \Bar{r}_{i,y}|}
    \right]
$$
This formulation allows for differentiation between compressive and tensile forces.

\subsection*{Gene Networks} Methods presented in Hiscock et al. \cite{hiscock2019} are adapted to model $N$ genes that can interact with each other, with the sign of the interaction term determining whether it is an activating or inhibitory coupling. Weights of the gene networks are shared by all cells, in the same spirit in which actual cells share the same genetic material. Gene networks are specified by the following ODE model:

\begin{gather}
    \frac{dg_i}{dt} = \phi(\Sigma_j W_{ij}g_j + b_i) + I_i - k_ig_i
\end{gather}
Considering the $i$-th gene, $g_i$ is the expression level, $W_{ij}$ is the regulatory influence of gene $j$ on $i$, $b_i$ determines the gene transcription rate in absence of regulation, $k_i$ is the degradation rate, and $I_i$ is a forcing signal coming from one of the sensed cell inputs. $\phi$ is a sigmoid function that ensures positive transcription rates saturating at high input levels. Local signals sensed by the cell (chemicals, stress, etc.) are fed as input signal to input nodes in the network (colored green in the diagrams in the main text). All other genes receive no external inputs. The last output gene's readout is the division propensity, and the last $N_\text{c}$ genes readout the secretion rates of chemicals. In the case where cellular adhesion is chemically regulated, there are also output genes that readout cadherin concentrations.

\section{The REINFORCE Optimization Algorithm}

\subsection{Underlying Theory}

REINFORCE is a widely used reinforcement learning algorithm that allows to optimize the action policy directly, in a context where stochastic decisions must be taken. This is the algorithm we use for most of our optimizations, and below are the main steps in its derivation.

\subsection*{Markov Decision Processes}
We assume that our simulation follows a Markov Decision Process (MDP). That is, the probability of a trajectory $\tau$ is the product of the probability of the initial state and the probabilities of each action and state transition along the trajectory. Let $\tau = (s_0, a_0, s_1, a_1, \ldots, s_T, a_T)$ represent a trajectory, where $s_t$ and $a_t$ are the state and action (cell division) at time $t$, respectively. Assume $\pi(a_t|s_t)$ is the policy (the division propensity calculated by the gene network), $P(s_{t+1}|s_t, a_t)$ is the state transition probability, and $P(s_0)$ is the initial state distribution. The probability of the trajectory $\tau$ is given by:

$$
P(\tau) = P(s_0) \prod_{t=0}^{T} \pi(a_t|s_t) P(s_{t+1}|s_t, a_t)
$$
\\
Where:
\begin{itemize}
\item[] $P(s_0)$ is the probability of the initial state $s_0$
\item[] $\pi(a_t|s_t)$ is the probability of taking action $a_t$ given state $s_t$
\item[] $P(s_{t+1}|s_t, a_t)$ is the probability of transitioning to state $s_{t+1}$ given state $s_t$ and action $a_t$.
\end{itemize}

\subsection*{Policy Gradients Theorem}
The goal is to maximize the expected return $J(\theta)$, where $\theta$ are the parameters of the policy $\pi_\theta(a|s)$. The expected return is defined as:

$$
J(\theta) = \mathbb{E}_{\pi_\theta}\left[G_t\right]
$$
where $G_t$ is the (possibly discounted) return starting from time step $t$. The expected gradient of the return with respect to the policy parameters $\theta$ can be computed using the log-derivative trick, also known as the REINFORCE algorithm. The expected return $J(\theta)$ is given by:

$$
J(\theta) = \sum_{\tau} P(\tau) G(\tau)
$$

The gradient of the expected return is:

$$
\nabla_\theta J = \nabla_\theta \sum_{\tau} P(\tau) G(\tau) = \sum_{\tau} G(\tau) \nabla_\theta P(\tau)
$$

Using the log-derivative trick, $\nabla_\theta P(\tau)$ can be written as:

$$
\nabla_\theta P(\tau) = P(\tau) \nabla_\theta \log P(\tau)
$$

Thus, the expected gradient is:

$$
\nabla_\theta J(\theta) = \sum_{\tau} P(\tau) G(\tau) \nabla_\theta \log P(\tau)
$$
\\
Using the definition of MDP given above, the gradient of the expected return with respect to $\theta$ can be simplified to only include the policy $\pi_\theta$:

$$
\nabla_\theta J(\theta) = \mathbb{E}_{\pi_\theta}\left[ \nabla_\theta \log \pi_\theta(a|s) G_t \right]
$$

\subsection*{Monte Carlo Estimation}
In practice, the expectation is estimated using Monte Carlo sampling. By sampling trajectories $\tau = (s_0, a_0, r_0, s_1, a_1, r_1, \ldots)$ from the policy $\pi_\theta$, the gradient can be approximated as:

$$
\nabla_\theta J(\theta) \approx \frac{1}{N} \sum_{i=1}^N \sum_{t=0}^{T-1} \nabla_\theta \log \pi_\theta(a_t^i|s_t^i) G_t^i
$$
where $N$ is the number of sampled trajectories, and $T$ is the length of each trajectory.

\subsection*{Parameters Update Rule}
The REINFORCE algorithm updates the policy parameters $\theta$ by gradient ascent (if we are considering the case of maximizing total rewards):

$$
\theta \leftarrow \theta + \eta \nabla_\theta J(\theta)
$$
where $\eta$ is the learning rate. In order to calculate the actual update, we substitute the Monte Carlo gradient estimate for the gradient derived above.

\subsection*{Baseline for Variance Reduction}
To reduce the variance of the gradient estimate, a baseline function $b(s_t) $ is introduced, leading to the modified gradient:

$$
\nabla_\theta J(\theta) = \mathbb{E}_{\pi_\theta}\left[ \nabla_\theta \log \pi_\theta(a|s) (G_t - b(s_t)) \right]
$$
\\
A common choice for the baseline is the state-value function $V^\pi(s_t)$, which represents the expected return from state $s_t$. Another common basic technique for variance reduction is to normalize the rewards after each batch of simulations, resulting in half of the actions being favored and half unfavored on average. We adopt the latter in this work.

\subsection{Optimization of Cell Divisions}

In our simulations, we consider the policy to be the gene regulatory network inside each cell. Cells sense inputs from the environment and use them to decide how likely they should be to divide next. Notice that a loss applied naively at the end of the simulation --- like for example a measure of distance from a given target --- would produce undefined gradients. Stochastic operations (like sampling a cell for division) do not have a mathematically well-defined differentiation rule; therefore as soon as the back-propagation procedure encounters such operations, the whole gradient becomes ill-defined (or zero, in practice in the code). In order to overcome this obstacle, we resort to techniques like the one presented in the previous section.

Our simulation presents an additional complication with respect to the policy gradient case defined above. Cells influence the environment not only through their division decisions but also, mainly, through the modulation of chemical secretion. This creates a long feedback loop connecting secretions at one step to the chemical landscape at the next step, which in turn influences division decisions. As a consequence, the gradient of a loss applied to cell division events -- like REINFORCE -- must be propagated through the environment updates till the very beginning of the simulation. This makes automatic differentiation algorithms critical for this sort of optimization.

In all our optimizations, we use the Adam optimizer for gradient descent (hyperparameters are listed below). Reported losses, here and in the main text, are validation losses calculated on a batch of simulations different from the one used for updating the parameters.

\clearpage
\section{Additional Results}

We report in this section on a set of additional experiments and results that we do not include in the main text.

\begin{figure*}
\begin{center}
\includegraphics[scale=.9]{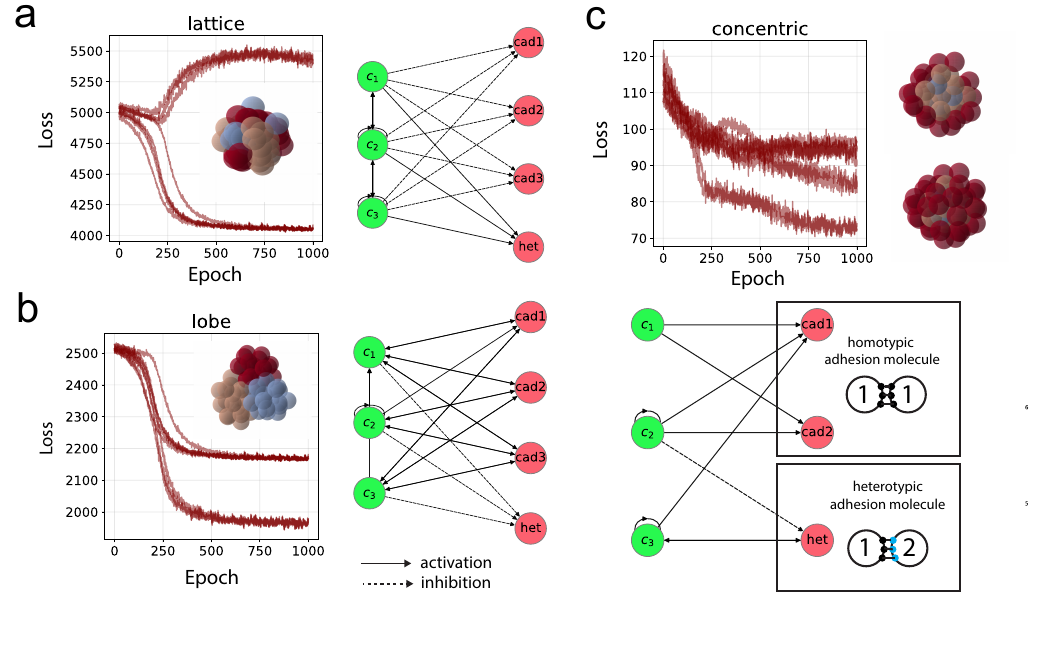}
\caption{\textbf{Chemical Regulation of Cellular Adhesion.} Learned chemical regulation of cellular adhesion guides the assembly of different spatial cluster arrangements. The top part of the panels show the training loss curves for three different optimization targets. The insets show the final state of a simulation after the learning, with colors denoting different cell types. The bottom sections show the learned and pruned gene network; \textit{Cad1}, \textit{Cad2}, and \textit{Cad3} are the homotypic cadherins of the three cell types, and \textit{Het} is the heterotypic cadherin expressed in all cell types. (a) Lattice-like structure. Cells are arranged to minimize the sum of pairwise distances between heterotypic cells. (b) Lobed structure. Cells minimize the sum of pairwise distances between homotypic cells. (c) Core-shell structure. Cells are arranged to minimize the distance of cells from a predefined radius, dependent on their type. The final configuration is here shown as a cross-section to highlight the internal structure.}
\label{fig:adhesion}
\end{center}
\end{figure*}

\subsection*{Chemical Regulation of Cellular Adhesion}
The differential adhesion hypothesis posits that populations of cells can adopt specific morphologies based on differences in adhesive strength \cite{steinberg_does_1970}. During morphogenesis, a cadherin code regulates the spatial organization of developing tissue \cite{tsai_adhesion_2020,bao_stem_2022}. Engineered cellular circuits can use gene regulatory networks to dynamically modulate adhesion and program self-organization \cite{toda_engineering_2020}. Here, we show that gene interactions can be learned that use chemical signaling to modulate homotypic and heterotypic adhesion to achieve a target spatial pattern in a population with different cell types.

In our model, a population of three cell types is subject to Brownian motion and interacts via a Morse pairwise potential. The well depth of the potential between a pair of cells is dictated by the concentration of homotypic and heterotypic “cadherins” in each cell. Each cell type produces its respective homotypic cadherin and a shared heterotypic cadherin, regulated by chemical signaling. Each cell type utilizes the same gene network, but can only secrete chemicals and express cadherins of its own type. To achieve a desired spatial organization, we construct three loss functions based on pairwise distances between cells.  For a lattice-like structure, we minimize the sum of pairwise distances between cells of different types (Fig \ref{fig:adhesion}a). To organize the structure into three lobes, we minimize the sum of pairwise distances between cells of the same type (Fig \ref{fig:adhesion}b). To achieve a core-shell structure, we minimize the distance of each cell to a pre-specified radius, based on the cell type (Fig \ref{fig:adhesion}c).

We model cell adhesion in 3D, with a cluster of 60 cells that do not divide. The cluster has three cell types, 3 cell-type specific signalling chemicals and 0 hidden genes. Each cell's gene regulatory network senses as input only local chemical concentrations, and outputs the concentration of a homotypic cadherin (controlling the well depth between the same cell type) and a heterotypic cadherin (controlling the well depth with a different cell type). 

The optimization process identifies effective solutions by allowing cells to observe their own type and regulate the pairwise potential between cell types. For a lattice-like structure, cells up-regulate their heterotypic cadherin and down-regulate their homotypic cadherin. Conversely, to create a lobed structure, they up-regulate their homotypic cadherin and down-regulate their heterotypic cadherin. These patterns of homo- and heterotypic adhesion are consistent with observations from cell culture sorting experiments \cite{tsai_adhesion_2020}. Although the formation of a lattice and lobe structure are trivial, achieving a core-shell structure with three cell types requires a more complex regulatory network to achieve the correct differential adhesion - the inner cells have strong homotypic adhesion, while the outermost cells require heterotypic adhesion to stick to the second layer of cells. (Fig \ref{fig:adhesion}c).

\begin{figure} 
\centering
\includegraphics{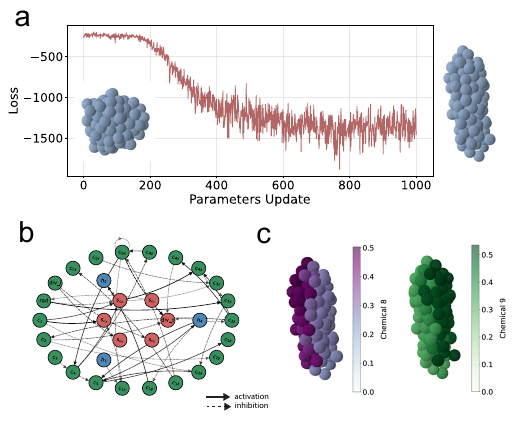}
\caption{\textbf{Elongation.} Maximization of the squared sum of the z-coordinates of cells, resulting in vertical elongation in the optimized cluster. (a) Loss curve for one successful optimization run. On the left the final configuration of a simulation with randomly initialized parameters, on the right the final state after learning. (b) Visualization of the optimized and pruned gene network, displaying activation and inhibition interactions that regulate cell behavior during elongation. (c) Two examples of emerging chemical gradients that are used to guide growth.}
\label{fig:elongation}
\end{figure}

\subsection*{Elongation (Single cell type)}We model the elongation of a single cell type in three dimensions, optimizing the cell cluster to minimize the sum of squared z-coordinates, thus promoting vertical elongation (Fig. \ref{fig:elongation}a).  This contrasts with the previous example by focusing on a single cell type and incorporating the cells' ability to sense chemical gradients.  Each cell's gene network receives as input local chemical concentrations, estimated chemical gradients, cell size, and division propensity (Fig. \ref{fig:elongation}b).  These inputs drive positional information inference through local physical communication. Simulations initiate with a single cell and proceed until the cluster reaches 120 cells.

Critically, cells autonomously establish or manipulate existing chemical gradients to achieve the global elongated shape, a feature not explicitly encoded in the optimization objective (Fig. \ref{fig:elongation}c).  The gene networks evolve to concentrate cell division primarily at the cluster's ends.  Because the cluster boundary dynamically shifts with cell division, the gene networks must implement a sophisticated control mechanism to sustain directed growth.

\begin{figure}
\centering
\includegraphics{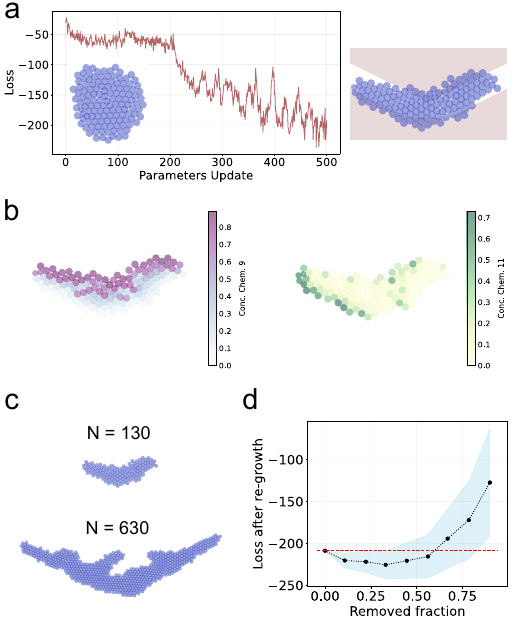}
\caption{\textbf{Branching.} Optimizing growth in a V-shaped reward landscape results in a learned branched structure of the final cluster. (a) Loss curve for one successful optimization run. On the left, the final configuration of a simulation with randomly initialized parameters. On the right, the final state after learning with a visualization of the reward mask (red areas are associated to penalties, white ones with rewards). (b) Two examples of emerging chemical gradients that are used to guide growth. (c) Final optimized state after 129 cell divisions, the same number used for the training (top) and with 500 extra cell divisions. Note the spontaneous branching on the main arms. (d) Loss of the final state after removal, at random, of a fraction of cells. The red dashed line is the loss associated to the final simulation state before cell removals.}
\label{fig:branching}
\end{figure}

\subsection*{Branching}
We simulate the growth of a cluster from a single cell to 130 cells, restricting ourselves to the 2-dimensional case for computational simplicity. The simulation is initialized with 1 cell in 2D, and run for 129 cell division steps to generate a cluster of 130 cells. Simulations are done with  20 chemicals and 50 internal genes. With a similar spirit to diffusion-based geometry \cite{iber2013}, we simulate permeable diffusive boundaries between the cell cluster and the surrounding implicit medium by connecting boundary cells to a ghost chemical sink node. Gene networks have access to the local chemical concentrations, estimates of local chemical gradients, their own cell radius and division propensity. They do not have access to their own absolute spatial coordinates. 

To learn more complex branched structures, we construct a position-based shape loss. At each time point, cell divisions that place cells within the region designating the desired shape are assigned a reward, the others a penalty. In this case, we impose a V-shaped spatial reward structure (Fig. \ref{fig:branching}a on the right) augmented with an asymmetry penalty on the x-coordinate, in order to promote the growth of both arms at the same time. Also in this case, a variety of chemical patterns emerge in the learned simulation (Fig  \ref{fig:branching}b), and the cluster learns to divide cells primarily at the tips of the branching structure. Clusters visually acquire a branched shape, even when simulated for 500 cell divisions more than optimized for. While shape errors clearly emerge, they do mostly in the form of smaller branches that begin to sprout from the main branched structure (Fig \ref{fig:branching}c). Furthermore, the learned network displays emergent regenerative capabilities. After reaching the final branched configuration, we remove a fraction of cells chosen at random and we allow the cluster grow back to its original size. As shown in Fig \ref{fig:branching}d, eliminating up to 60\% of cells still results in the regeneration of a branched structure. Even more interestingly, it appears that randomly removing cell actually promotes error-correction in the regrowth, reaching even slightly better performances in the training loss.

\clearpage
\section{Experiment Details}
\subsection*{Chemical Regulation of Cellular Adhesion} The simulation is initialized with 60 cells and 3 cell types, and done in 3D. Cell positions and identities are randomly initialized. For the lattice and lobe structures, the cluster consists of equal numbers of each cell type (20 cells each); for optimizing a core-shell structure, the cluster consists of 10 blue cells, 20 yellow cells and 30 red cells. The simulation is run for 50 time steps, with 200 Brownian relaxation steps between each time step; the temperature is high to allow for sufficient rearrangement of the system without trapping. No cell division occurs. The simulation contains 0 internal genes, and each cell type can secrete 3 cell type specific morphogens. 
\\
\\
\textit{Cadherin.} The system contains $N_{\text{ctype}}$+ 1 cadherins, with each cell able to regulate concentrations of its cell type specific cadherin (\textit{cad1}, \textit{cad2}, etc.), as well as a heterotypic cadherin (\textit{het}). The gene network takes as input sensed chemicals and regulates only the concentrations of cadherins. The well depth of the pairwise potential between cells is calculated as the sum of the concentration of homotypic cadherins (if they are the same cell type) or the sum of the concentration of heterotypic cadherins (if they are different cell types). This also ensures that the pairwise well depth matrix is always symmetric. The calculated well depth value is then scaled by a sigmoid to be between .8 and 3.8, ensuring a stable range of values are used for the Morse potential.  
\\
\\
\textit{Optimization.} Since there is no cell division in this simulation, we directly backpropagate through the simulation without using REINFORCE. To improve the quality of gradients when backpropagating through long simulations, we "discount" gradients in the backward pass by a discounting factor of 0.99 - this results in gradients from time steps closer to the end of the simulation providing a greater contribution to the optimization. Training is done over a batch of 4 simulations, and the validation loss is averaged over a batch of 64 simulations. Additionally, at each gradient descent step, we constrain the learned network by setting to zero all outgoing edges from output nodes. 
\\
Three different loss functions are utilized. To learn a lobe-like arrangement, the following loss is used:

$$
\frac{1}{N_{t}}\Sigma_t\Sigma_{\text{cell} \, i}\Sigma_{\text{cell} \, j} \begin{cases}
    \text{dist}(i,j), & \text{if}\ \text{type}(i) == \text{type}(j) \\
    0, & \text{otherwise}
\end{cases}
$$
This is the sum of pairwise distances between cells of the same type, averaged over all time steps of the simulation. Conversely, to attain a lattice-like structure,  the sum of pairwise dustances between unlike cell types is minimized:

$$
\frac{1}{N_t}\Sigma_t\Sigma_{\text{cell} \, i}\Sigma_{\text{cell} \, j} \begin{cases}
    0, & \text{if}\ \text{type}(i) == \text{type}(j) \\
    \text{dist}(i,j), & \text{otherwise}
\end{cases}
$$
For the core-shell structure, the distance of cells from a predefined radius (based on the cell's type) with respect to the center of the core cells is minimized:

$$
C = \frac{1}{N_1}\Sigma_{i \in \text{type 1 cells}} \, \Bar{r_i}
$$

$$
\\
\frac{1}{N_t}\Sigma_t\Sigma_{\text{cell} \, i} \begin{cases}
    (\text{dist}(i, C) - 1)^2, & \text{if}\ \text{type}(i) == 1 \\
    (\text{dist}(i, C) - 2)^2, & \text{if}\ \text{type}(i) == 2 \\
    (\text{dist}(i, C) - 3)^2, & \text{if}\ \text{type}(i) == 3
\end{cases}
$$
Here, C is the average position of the core cells, $N_1$is the number or core cells, type($i$) is the cell type of cell $i$, and dist($i$,$j$) is the Euclidean distance between the positions of cells $i$and $j$. To optimize, we used a learning rate of 0.001 for 1000 gradient descent steps with Adam optimizer. The weights of the gene network are not regularized, but the network is distilled by pruning.

\subsection*{Elongation (Two cell types)}
This is the elongation example shown in the main text. The initial state is composed of 15 cells, 10 source cells and 5 proliferating cells. The simulation is run for 135 cell division events, so the final state has a total of 150 cells. Source cells secrete the chemical factor that diffuses through space. Proliferating cells sense the chemical factor and use this information to modulate cell division propensity.

\textit{Optimization.} The loss function is calculated as the sum of the squares of the x-coordinate of the cells. Training is done on a batch of 4 simulations. We use a learning rate of 0.005 for 1000 gradient descent steps with Adam optimizer. The weights of the learned gene network are L1-regularized with $\lambda=1.$.

\subsection*{Elongation (Single cell type)}
The forward simulation is initialized with 1 cell in 3D, and run for 119 cell division steps. The simulation consists of 10 chemicals and 32 internal genes, and the gene network takes as input sensed chemical concentrations, chemical gradients, division propensities, and cell radii. Diffusion occurs in a closed system.
\\
\\
\textit{Optimization.} The loss function is calculated as the sum of the squares of the z-coordinate of each cell; minimizing this forces the cluster to elongate vertically. Training is done on a batch of 4 simulations, and a batch of 4 simulations is used to evaluate the validation loss. We use a learning rate of 0.001 for 500 gradient descent steps with Adam optimizer. The weights of the learned gene network are L1-regularized with $\lambda=.1$. The weights of the gene network are initialized to values drawn from a normal distribution with $\sigma=.1$. The weights of the learned gene interaction matrix are pruned (set to zero) if they are below $\epsilon = 0.9$. This threshold is determined by pruning the learned parameters at various thresholds and measuring the loss of 50 simulations run with pruned parameters (Fig. \ref{fig:elo_bra_hom}a). 

\subsection*{Branching}
\textit{Optimization.} The loss function is calculated by defining a shape mask that specifies the region of space cells should occupy to be rewarded. Cells that are within the mask are given a reward of 3.0 and cells that are not are given a penalty of -1. These values are summed over all cells and an asymmetry penalty is added to prevent cells from forming just one branch. The asymmetry penalty is given by the absolute value of the sum of the x-coordinates of cells. We use a learning rate of 0.01 for 500 gradient descent steps with Adam optimizer. The weights of the learned gene network are L1-regularized with $\lambda=.1$. The weights of the gene network are initialized to values drawn from a normal distribution with $\sigma=0.1$. 
\\
\\
\textit{Regeneration.}
To obtain the regeneration plot shown in Fig \ref{fig:branching}d, forward simulations are run using optimized parameters. A fraction $f$ of cells is removed  by deleting $f*N$ random indices, where $N$ is the total number of cells in the structure, and a simulation is run for $(1 - f)*N$ steps using the partial structure as the initial state. The final loss after regeneration is averaged for 15 simulations, for various $f \in [0,1]$.

\subsection*{Chemical Homeostasis}

The simulation is initialized with 20 cells, and run for 100 cell division steps. The initial cells are divided into two cell types, where each cell type can only secrete one of two chemicals in the simulation. An initial imbalance between the two cell types is set by assigning 20\% of the initial cells to be the first cell type and the remaining to be the second cell type. The simulation contains 8 internal genes, and the gene network takes as input the sensed chemical concentrations. Diffusion occurs in a closed system.
\\
\\
In this case each cell type is equipped with a different regulatory network, shared among all cells of the same type. Each cell type can still only secrete one chemical of the two but sense the concentration of both of them. Fig \ref{fig:chemhomeo_twonetworks}a showcases gene networks for cell type 1 and 2 obtained from different optimization runs. Out of 20 optimization runs, 12 converged to the architectures in the top row, the other on one of the other two cases. Notice the consistency in the discovered architectures, especially of the stronger (bolder) connections. Fig \ref{fig:chemhomeo_twonetworks}b schematically represents the model optimized in this context.
\\
\\
\textit{Optimization.}
The loss function is calculated as the imbalance in the number of cell types, given by:

$$
\mathcal{L}_t = |\text{num type 1 cells} -  \text{num type 2 cells}|
$$

We used a learning rate of 0.01 for 400 gradient descent steps with Adam optimizer. The weights of the learned gene network are L1-regularized with $\lambda=.1$. To have the untrained state maintain the initial imbalance of the cell types in the generated structure, the weights of the gene network are initialized to very low values, resulting in a homogeneous division propensity that could only be broken by learning how to chemically regulate division to maintain equal numbers of cell types.

\begin{figure}
\centering
\includegraphics[width=\textwidth]{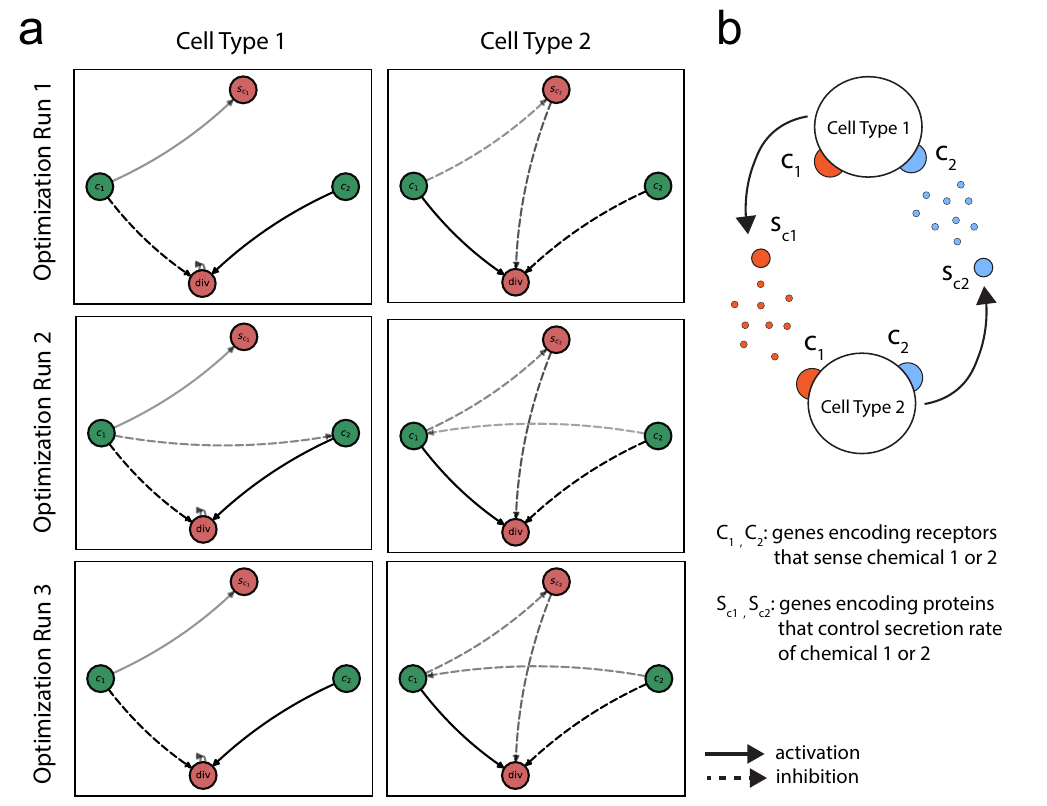}
\caption{\textbf{Chemical Regulation of Homeostasis with Cell Type Specific Networks}. (a) Gene network architectures for each cell type (on columns) obtained from different optimization runs (on rows). Green nodes are sensory inputs (i.e. sensed chemical concentrations, while red nodes are outputs (secretion of the cell type specific chemical and division probability). (b) Schematic representation of the model we optimize over.}
\end{figure}
\label{fig:chemhomeo_twonetworks}
\FloatBarrier

\begin{figure}
\centering
\includegraphics[width=\textwidth]{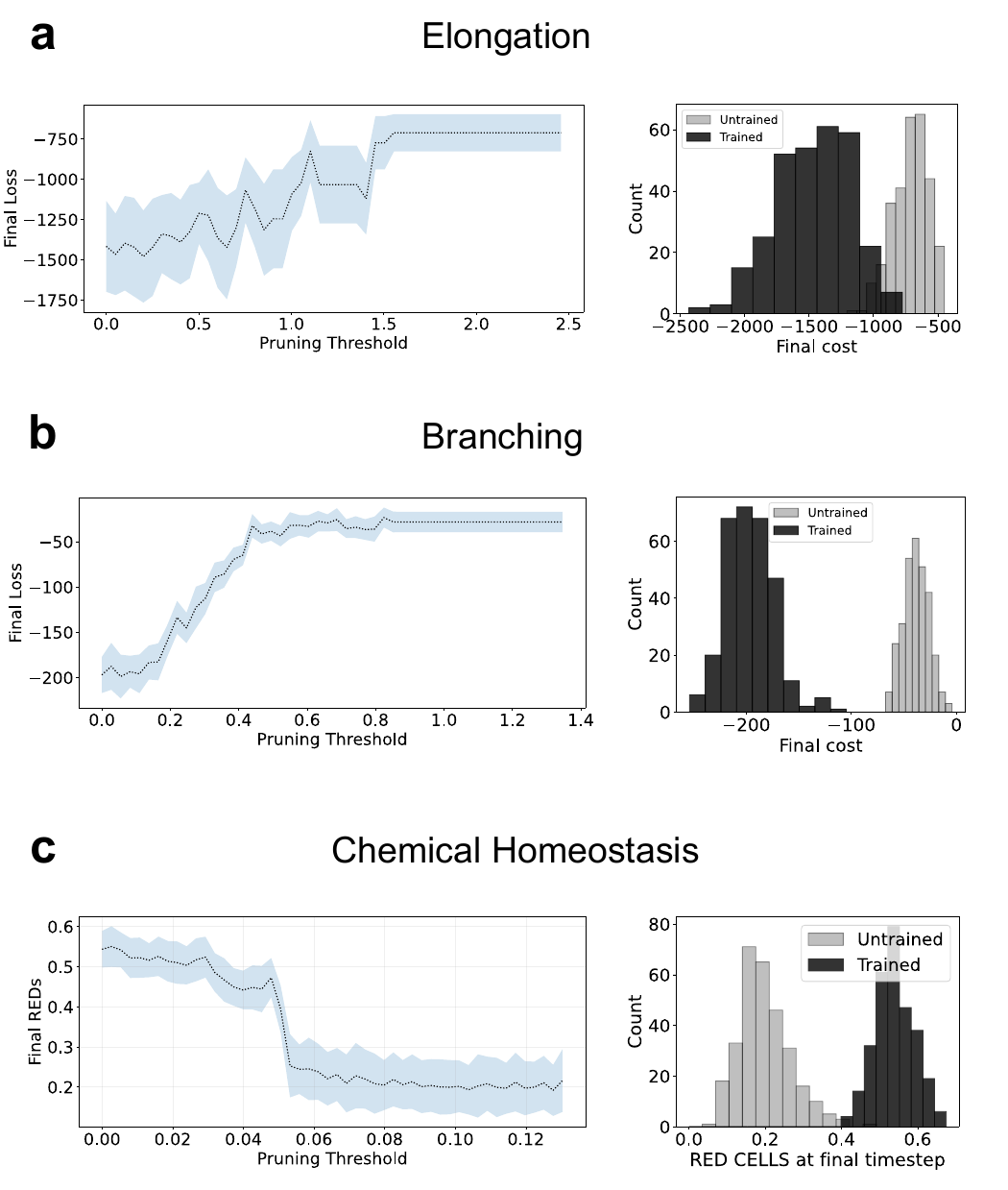}
\caption{\textbf{Elongation, Branching and Chemical Regulation of Homeostasis}. Final losses for a range of pruning thresholds shown on the left for gene network learned for (a) Elongation, (b) Branching, and (c) Chemical Homeostasis. Losses are averaged over 50 simulations, and shaded area shows one standard deviation. On the right, histograms comparing the distribution of final losses from the untrained and trained models for each of the three cases; there is a clear separation of the trained model from the untrained.}
\end{figure}
\label{fig:elo_bra_hom}
\FloatBarrier

\subsection*{Mechanical Control of Cell Proliferation} 
\begin{figure}
\centering
\includegraphics{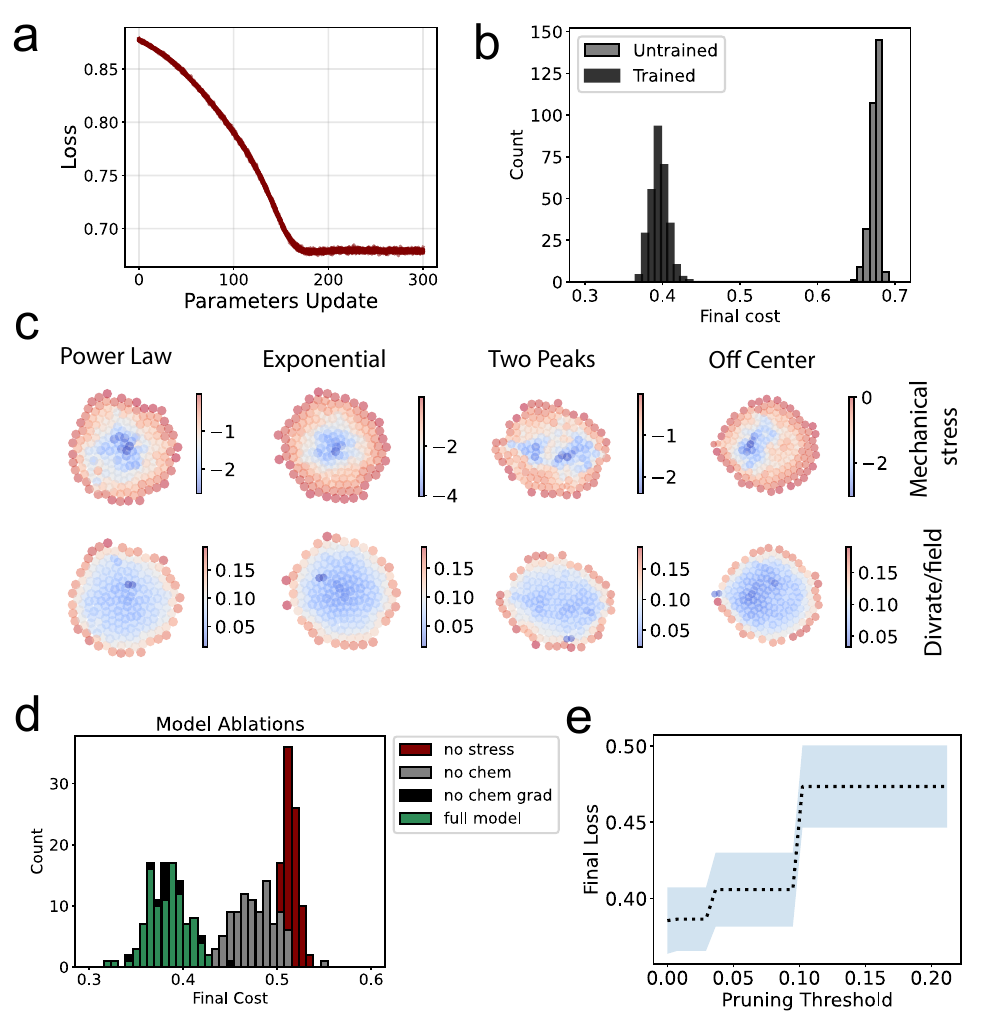}
\caption{\textbf{Mechanical Control of Cell Proliferation.} (a) Final costs of 300 forward simulations ran with trained parameters and untrained parameters. (b) Final costs of 300 forward simulations ran with trained parameters, ablating one of the input features to the gene network. Ablation is performed by setting outgoing edges from the desired feature to zero. Ablations show that mechanical stress is the most important feature used by the gene network to regulate division rates. (c) Visualizations of forward simulations run with trained parameters, but with a different shape of field than the one trained on. Top row shows the mechanical stress accumulation under different kinds of field, which follows the prescribed field. Bottom row shows the division rate unscaled by the field - the learned values weakly follow the pattern of stress accumulation. (d) Pruning plot showing final costs for various pruning thresholds of network weights.  }
\label{fig:homgrowth}
\end{figure}

The forward simulation starts from one cell and is run for 200 time steps. Throughout the simulation, an external field is established over the cells to mimic the effect of a growth factor gradient. The value of the field scales as $1/r^2$, with $r$ being the distance to the center of the cluster. At every cell division step, the division propensity calculated by the gene network is scaled by the value of this field:

$$d_i = d_i*\frac{1}{2 + 0.4*r^2}$$
\\
where $d_i$ is the division propensity of cell $i$.  This emulates the effect of a growth factor gradient - cells closer to the center are more likely to divide than cells closer to the edge. The simulation is run with 2 chemicals and 4 hidden genes, with the gene network being able to sense the cell's chemical concentration, chemical gradients, and mechanical stress. Mechanical relaxation occurs for a fewer number of steps in this simulation than in the other examples to prevent rearrangements from dissipating stress. Diffusion occurs with heuristic boundary conditions.
\\
\\
\textit{Optimization.}
Optimization is performed with a batch of 4 simulations for training, and a batch of 16 simulations for evaluation of the validation loss. We use a learning rate of 0.001 for 300 gradient descent steps with Adam optimizer. The weights of the learned gene network are L1-regularized with $\lambda=.1$. The loss function is the coefficient of variation of division propensities at the final time step. This optimization worked best without a REINFORCE gradient estimate - information from the last time step is enough to learn a mechanism. The weights of the gene network are initialized to zero so that the initial cluster grows as prescribed by the growth factor gradient.
\\ To evaluate the nature of the learned mechanism, inputs to the trained model are ablated, by replacing the weights of the gene network corresponding to a specific input to zeros. In this way, we ablated chemicals, chemical gradients and stress as inputs, finding that ablating stress results in markedly decreased performance of the mechanism (Fig \ref{fig:homgrowth}c). We additionally tested the learned mechanism on other kinds of fields (as described in the main text) and saw that the distribution of unscaled division propensities seem to match the distribution of stress buildup in the cluster (Fig \ref{fig:homgrowth}d).
\\

\FloatBarrier

\clearpage
\section{Scaling of Optimization Performance with Model Complexity}

\begin{figure*}[h!]
\begin{center}
\includegraphics[width=\textwidth]
{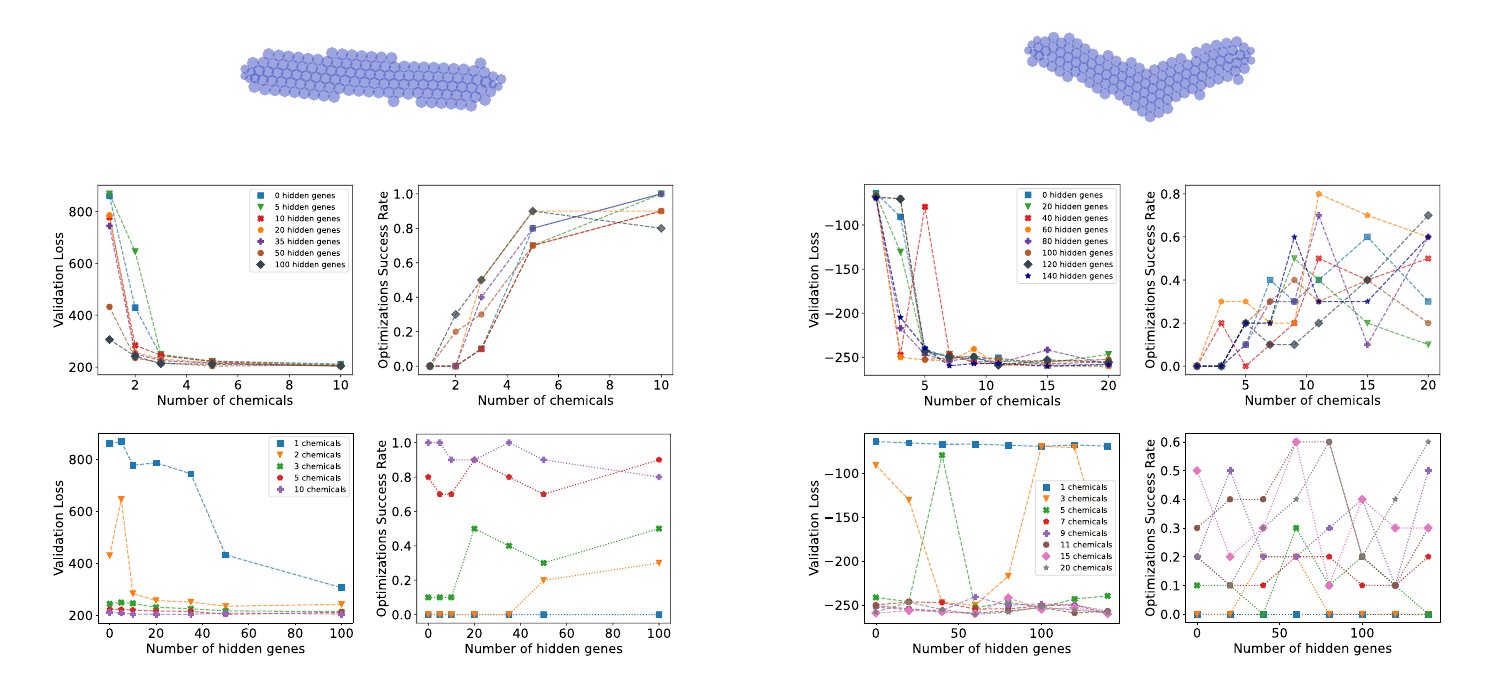}
\caption{\textbf{Scaling of Optimization Performance with Model Complexity.} (a) Scaling of validation loss and success rates for elongation. Losses are collected for 10 optimizations run from a different random seed for each condition. Training is done over 500 gradient descent steps and a batch of 8 simulations. (b) Scaling plots for V-shape case. Losses are collected for 10 optimizations run from a different random seed for each condition. Training is done over 300 gradient descent steps and a batch of 8 simulations.}
\label{fig:scaling}
\end{center}
\end{figure*}

Over-parametrized deep networks have shown a remarkable generalizability in learning target functions. In a similar vein, we preliminarily investigated how our optimization performance scales with the complexity of the regulatory network in our cells -- i.e, the number of secreted morphogens and the number of interacting genes (Fig \ref{fig:scaling}). We restrict ourselves for simplicity to the 2D case of growing elongated and branched structures. In learning the decision-making circuits required for cells to grow into a pre-specified shape (either an elongated cluster or a branched "V-shaped" structure), a lower validation loss is reached by increasing the number of morphogens each cell can locally communicate with. The rate of success of optimizations - the fraction of optimizations that are able to successfully create the desired shape - also increases with the number of morphogens. Without a sufficient number of morphogens, however, none of the optimizations can learn a successful decision-making network to create the desired shape. In the case of the V-shape, increasing the number of interacting genes did not have a pronounced effect. Interestingly, for the case of the elongated cluster, increasing the number of genes helped learn successful mechanisms in systems with low numbers of morphogens. In both cases, a sufficient number of both morphogens and genes is necessary to learn a successful mechanism. These findings indicate a minimal complexity required to build shapes and an interesting tradeoff between the quantity of communication signals (number of morphogens) versus the complexity of the local communication network (number of interacting genes).

\FloatBarrier

\begin{table}\centering
\caption{Definition of Notation}

\begin{tabular}{lrr}
\textbf{Notation} & \textbf{Definition} \\
\midrule

T & number of timesteps \\
N & number of cells in final state \\
n & number of cells in initial state \\
Dim & dimension of cell property \\
R & cell radii \\
$N_c$& number of chemicals in system \\
$N_{\text{ctype}}$& number of celltypes in system \\
$N_t$& number of timesteps \\
$\Bar{r_i}$& position of cell $i$\\
$d_i$& division propensity of cell $i$\\
$\Bar{g_i}$& gene concentration vector of cell $i$\\
$c_k (\Bar{r}_i)$& chemical $k$concentration at position $\Bar{r}_i$\\ 
D & diffusion coefficients \\
S & secretion rates \\
K, $k_i$& degradation rates \\
$W_{ij}$& interaction weight of gene $j$ regulating gene $i$\\
$I_i$& input signal to gene $i$\\
\bottomrule
\end{tabular}
\end{table}

\FloatBarrier

\bibliography{refs}